\documentclass[a4paper,11pt]{article}
\usepackage{jheppub} 
\usepackage[T1]{fontenc}
\title{ Super-Resonant Dark Matter}
\author[1]{Csaba Cs\'aki,}\emailAdd{csaki@cornell.edu}
\author[1]{Andrew Gomes,}\emailAdd{awg76@cornell.edu}
\author[2]{Yonit Hochberg,}\emailAdd{yonit.hochberg@mail.huji.ac.il}
\author[2]{Eric Kuflik,}\emailAdd{eric.kuflik@mail.huji.ac.il}
\author[3]{Kevin Langhoff,}\emailAdd{klanghoff@berkeley.edu}
\author[3,4,5]{Hitoshi Murayama\astfootnote{Hamamatsu Professor}}\emailAdd{hitoshi@berkeley.edu}
\affiliation[1]{Department of Physics, LEPP, Cornell University, Ithaca, NY 14853, USA}
\affiliation[2]{Racah Institute of Physics, Hebrew University of Jerusalem, Jerusalem 91904, Israel}
\affiliation[3]{Berkeley Center for Theoretical Physics, Department of Physics, University of California,\\ Berkeley, CA 94720, USA}
\affiliation[4]{Theoretical Physics Group, Lawrence Berkeley National Laboratory, Berkeley, CA 94720, USA}
\affiliation[5]{Kavli IPMU (WPI), UTIAS, The University of Tokyo, Kashiwa, Chiba 277-8583, Japan}
\date{ \today }
\renewcommand{\thefootnote}{\arabic{footnote}}

\usepackage{braket,graphicx,float,physics,amsthm,amssymb,amsfonts,xcolor,subfiles,indentfirst,booktabs,subcaption,slashed,tikz-feynman,comment}

\newcommand{\astfootnote}[1]{%
\let\oldthefootnote=\thefootnote%
\setcounter{footnote}{0}%
\renewcommand{\thefootnote}{\fnsymbol{footnote}}%
\footnote{#1}%
\let\thefootnote=\oldthefootnote%
}

\usepackage[export]{adjustbox}

\graphicspath{{Images/}}

\newcommand{\Mes}{\widetilde{M}}
\newcommand{\Mesd}{{{\Mes}^\dagger}}
\newcommand{\Holo}{\Lambda_\text{H}}

\newcommand{\beq}{\begin{equation}}
\newcommand{\eeq}{\end{equation}}
\newcommand{\ba}{\begin{array}}
\newcommand{\ea}{\end{array}}
\newcommand{\bea}{\begin{eqnarray}}
\newcommand{\eea}{\end{eqnarray} }
\newcommand{\be}{\begin{eqnarray}}
\newcommand{\ee}{\end{eqnarray} }
\newcommand{\bal}{\begin{align}}
\newcommand{\eal}{\end{align}}
\newcommand{\bi}{\begin{itemize}}
\newcommand{\ei}{\end{itemize}}
\newcommand{\ben}{\begin{enumerate}}
\newcommand{\een}{\end{enumerate}}
\newcommand{\bc}{\begin{center}}
\newcommand{\ec}{\end{center}}
\newcommand{\bt}{\begin{table}}
\newcommand{\et}{\end{table}}
\newcommand{\btb}{\begin{tabular}}
\newcommand{\etb}{\end{tabular}}

	\abstract{We introduce \textit{Super-Resonant Dark Matter}, a model of self-interacting dark matter based on the low energy effective theory of supersymmetric QCD. The structure of the theory ensures a resonant enhancement of the self-interactions of the low energy mesons, since their mass ratio is set by the number of colors and flavors. The velocity dependence of the resonantly enhanced self-interactions allows such theories to accommodate  puzzles in small scale structure that arise from dark matter halos of different sizes. The dark matter mass is then predicted to be around 3-4 MeV, with its abundance set by freeze-in via a kinetically mixed dark photon.   

}

\begin{document}
\maketitle

\section{Introduction}

Dark matter (DM) is a major driving force in the fields of particle physics, cosmology and astrophysics, and is potentially one of the best clues to physics beyond the Standard Model (SM). The $\Lambda $CDM model of cosmology is currently the best model we have for explaining many cosmological experiments involving big-bang nucleosynthesis (BBN), the cosmic microwave background (CMB) and large scale structure (LSS), which occur at scales of $\mathcal{O}(1\text{ Mpc}) $ and larger. However, several observations of small scale structure are hard to explain with collisionless cold DM. Some of the most studied small scale puzzles are  known as `core vs. cusp' \cite{CoreVsCusp1,CoreVsCusp2,CoreVsCusp3,CoreVsCusp4}, 
  `missing satellites' \cite{MissingSatelites1,MissingSatelites2}, 
 `too big to fail' \cite{TooBigToFail1,TooBigToFail2}, and the `diversity of rotation curves' \cite{DiversityOfRotations}. Several proposed solutions to these small scale structure puzzles exist, such as warm dark matter and the inclusion of dissipative baryonic processes into the numerical simulations \cite{Baryons1996,Baryons1999,Baryons2001,Baryons2002}. One such compelling notion is self-interacting dark matter (SIDM): it has been known for some time that if dark matter self-interacts, these self-interactions can potentially solve the small scale puzzles \cite{Spergel_2000,DiversitySIDM,SimulationsSIDM1,SimulationsSIDM2}.

Importantly, these small scale puzzles occur at several different length scales, or effectively, several different velocity scales. If one assumes that the DM self-interactions are velocity independent, then it is difficult to explain all small scale structure puzzles with SIDM while not violating experimental upper limits of the self-interaction strength. An alternative path is to invoke moderate velocity dependence in the self-interaction cross section \cite{Kaplinghat:2015aga}. This allows the accommodation of constraints on DM self-interactions set at relative velocities of $\mathcal{O}(10^3-10^4 \text{ km/s})$ while still allowing the self-interactions to be strong enough to explain the small scale puzzles which have velocities of $\mathcal{O}(10^2 - 10^3 \text{ km/s})$.

A simple method for producing velocity dependent cross sections is through an $s$-channel resonance \cite{VelocityDependence2019}. One mechanism for producing these resonances naturally involves dark mesons that are composed of combinations of light and heavy dark quarks. The process of interest is then analogous to the $\Upsilon(4s)-B\bar{B}$ resonance exploited in $B-$factory colliders~\cite{VelocityDependenceHQ2020}.

In this paper we introduce a supersymmetric SIDM model, which we call \textit{Super-Resonant Dark Matter} (SRDM). The main features of the model are:
\begin{enumerate}
    \item {\bf SUSY}: A supersymmetric dark sector that is compatible with the Minimal Supersymmetric Standard Model (MSSM) and is UV complete in the sense of introducing no naturalness problems. SRDM is assumed to couple to the MSSM via a dark photon.
    \item {\bf SQCD}: The UV description of the dark sector is supersymmetric QCD (SQCD), while the low-energy effective theory is one of mesons with an Affleck-Dine-Seiberg (ADS) superpotential \cite{ADS}. Degenerate supersymmetric quark masses, enforced via flavor symmetry, are introduced to stabilize the potential.
    \item {\bf Resonance}: Like non-SUSY QCD, the model experiences chiral symmetry breaking. The extra control over the dynamics that is provided by SUSY ensures that the flavor singlet and adjoint mesons come in a 2:1 mass ratio at tree level. This is a new mechanism for naturally producing resonances.
    \item {\bf Splitting}: To account for small scale structure puzzles at different velocities the above mass ratio must be shifted by an $\mathcal{O}(10^{-7})$ amount. In SRDM, this extremely small mass ratio splitting is achieved naturally by 1-loop corrections from meson self-interactions.
    \item {\bf Predictive}: Resonant SIDM models generally have three continuous parameters which determine self-interactions and need to be fixed, giving them flat directions in $\chi^2-$fits. For a fixed number of flavors, our model has only two and picks out a specific dark matter mass and self-interaction strength.
    \item {\bf Charged to Neutral Ratio}: The relic DM is produced via freeze-in. It can be completely neutral under the dark $U(1)$ or can have an $\mathcal{O}(1)$ fraction of charged DM, depending on the value of the dark photon coupling.
\end{enumerate}

The structure of the paper is as follows. In Section \ref{section:Model} we describe the details of the model and how the crucial mass ratio is generated via chiral symmetry breaking. We also describe how the model is coupled to a spontaneously broken $U(1)_D$ gauge symmetry, and how this massive dark photon is coupled in turn to the Standard Model. Additionally, we analyze the effect of supersymmetry breaking. In Section \ref{section:Self Interactions} we study the DM self-interactions and determine the parameter space which could allow our SRDM model to account for certain small scale structure puzzles. In Section \ref{section:Relic Abundance} we  calculate the relic abundance and determine through which mechanisms the correct relic abundance can be achieved. In Section \ref{section:Conclusion} we summarize our results and discuss possible ways of extending the analysis of the paper. We also include in Appendix~\ref{section:masscalc} a calculation of the 1-loop mass corrections and in Appendix~\ref{anom_match} we discuss anomaly matching as it relates to dark meson stability.

\section{Model}
\label{section:Model}
\subsection{Model overview}
\label{subsection:Model overview}

The model consists of a visible sector coupled weakly to a dark sector via a dark photon. The visible sector is identified with the Minimal Supersymmetric Standard Model (MSSM), into which our SM is embedded. The scale of superpartner masses is taken sufficiently large such that all direct collider bounds are evaded (at the price of some sub-percent level fine tuning). We assume that a supersymmetry (SUSY) breaking sector is coupled indirectly to the visible sector, giving rise to the soft breaking mass terms of the MSSM. Importantly, we assume that the only coupling of the dark sector to SUSY breaking is via the MSSM, leading to an almost completely supersymmetric dark sector, with small calculable SUSY breaking effects that we explain in Section \ref{subsection:Effect of SUSY breaking}. This will allow us to reliably predict the spectrum of the dark sector in a theory that has underlying strong interactions. It will also make degeneracies in the spectrum natural. Note that the mass scales we find in the dark sector are small compared to the scale of SUSY breaking in the MSSM.

For the dark sector, we simply use supersymmetric quantum chromodynamics (SQCD) with $N_c$ colors and $N_f$ flavors. In the UV, the matter content is just $N_f$ copies of quark and anti-quark chiral superfields transforming in the fundamental representation of the $SU(N_c)$ non-abelian gauge group. The quark masses $m_q$ are assumed to be degenerate and below the confinement scale, $m_q \ll \Lambda$. The light degrees of freedom in SUSY gauge theories always correspond to holomorphic gauge invariants (gauge invariants formed without the use of complex conjugation)---in the case of SQCD, these will be the meson (and when $N_c \leq N_f$, baryon) superfields\footnote{We will generally use the term \textit{meson} to mean the meson superfield, not just the scalar component. Similarly with the term \textit{quark}.}. When $N_c \leq N_f$ these composite fields are interpreted as actual confined degrees of freedom, while when $N_c > N_f$ (this work) the gauge group is partial broken and the mesons are the remaining uneaten massless degrees of freedom describing the ``moduli space''. The couplings between these mesons is non-perturbative at the confinement scale but become calculable at scales below the confinement scale, as we will explain in detail below.

\subsection{The supersymmetric spectrum of resonant models}
\label{subsection:The supersymmetric spectrum of resonant models }

First we give a brief overview of the supersymmetric spectrum of our hidden sector based on quarks and anti-quarks with $SU(N_f)_L \times SU(N_f)_R$ flavor symmetry in the fundamental (and anti-fundamental) of $SU(N_c)$, with masses below the strong coupling scale. As explored in the 1990's, the detailed form of the low-energy dynamics of these theories depends on the exact relation between the number of flavors $N_f$ and the number of colors $N_c$. For $N_f< N_c$, non-perturbative dynamics gives rise to an Affleck-Dine-Seiberg (ADS) type superpotential
\begin{align}
W_\text{ADS} =& \left(N_c-N_f\right) \left(\frac{\Holo^{3N_c-N_f}}{\det \Mes} \right)^{\frac{1}{N_c-N_f}}\label{eq:W_ADS} \\
\Mes_{ij} =& Q_i^a \bar{Q}_{a\, j} \label{meson_quarks}
\end{align}
where $\Holo$ is the holomorphic dynamical scale\footnote{This is the scale dynamically generated by the running of the holomorphic gauge coupling, which appears in front of the gauge kinetic term and whose beta function is 1-loop exact. The true dynamical scale $\Lambda$ is obtained by canonically normalizing the gauge kinetic term, and is given by the NSVZ beta function. The two differ by an $\mathcal{O}(1)$ irrelevant for our purposes and we will interchange them freely.}, $\Mes$ denotes the $N_f \times N_f$ meson matrix,  $Q$ and $\bar{Q}$ are the underlying chiral quark superfields, and $a$ is a color index while $i,j$ are flavor indices. Note that $\Mes$ has (non-canonical) dimension $2$. In the absence of other terms, the dependence of the potential on inverse powers of the meson field implies the expectation values of the fields will run off to infinity.

In order to stabilize the potential, we introduce the quark mass term
\begin{align}
    W_\text{mass}=m_q^{ij} Q_i\bar{Q}_j = \Tr m_q \Mes
\end{align}
 As the quark mass increases towards the dynamical scale, the meson VEV decreases towards the dynamical scale thereby increasing the gauge coupling at the point where the gauge group is broken.

Just like the quark mass term in the superpotential, the quark kinetic terms in the K\"{a}hler potential can also be written in terms of the meson fields. At tree level it is given by \cite{Affleck:1984xz} substituting the gauge field $V$ equations of motion into the quark K\"{a}hler potential $Q^\dagger e^V Q + \bar{Q}^\dagger e^{-V^T} \bar{Q}$ and projecting onto the meson field
\begin{align}\label{kahler_tree}
K_\text{tree} = 2 \Tr \sqrt{\Mes^\dagger \Mes} \ .
\end{align}
We will work throughout with the quark mass far below the confinement scale. The squark VEVs are therefore large and hence break the gauge group well above the confinement scale. Being at weak coupling, Eq.~(\ref{kahler_tree}) is an excellent approximation.

In the following, we will decompose the meson matrix into its singlet and adjoint components
\begin{align}\label{mes_def}
\widetilde{M} = \frac{1}{\sqrt{N_f}} S \cdot \mathbf{1} + M
\end{align}
where $\Tr M=0$. Taking a supersymmetric (flavor universal) quark mass matrix $m_q^{ij}=\mu \delta^{ij}$,\footnote{For readers ideologically inclined against global symmetries, one can very weakly gauge $SU(N_f/2)\times SU(N_f/2) \times \mathbb{Z}_2$ to ensure the flavor universality, as long as the gauge coupling is small enough to avoid the annihilation of dark matter pions into gauge bosons $\alpha \ll 10^{-5}$ which does not contradict the weak gravity conjecture.} our theory takes the form
\begin{align}
K =& 2 \Tr \sqrt{\Mes^\dagger \Mes} \\
W =& (N_c-N_f)\Big(\frac{\Lambda^{3N_c-N_f}}{\det \widetilde{M}}\Big)^{\frac{1}{N_c-N_f}} + \sqrt{N_f}\mu S
\end{align}

We first consider the K\"{a}hler potential.  Since we know that the superpotential will give the singlet a VEV, let us implement this with the shift $\Mes \rightarrow v^2 + \Mes$, or equivalently $S\rightarrow \sqrt{N_f}v^2 + S$. Expanding around this VEV gives
\begin{align}
\begin{split}
K &= 2 \Tr \sqrt{\Mes^\dagger \Mes} \\
&= \frac{1}{2v^2} \Tr\Mes \Mesd - \frac{1}{8v^4} \left( \Tr \Mes^2 \Mesd + \Tr \Mes \Mes^{\dagger 2} \right) + \dotsb
\end{split}
\end{align}
where we have discarded the sub-quadratic terms that do not contribute to the Lagrangian. Rescaling the mesons $\Mes \rightarrow \sqrt{2}\,v\Mes$ to get canonically normalized fields leaves us with
\begin{align}
\begin{split}\label{kahler11}
K &= \Tr\Mes \Mesd - \frac{1}{2\sqrt{2}\,v} \left( \Tr \Mes^2 \Mesd + \Tr \Mes \Mes^{\dagger 2} \right) + \dotsb
\end{split}
\end{align}

We now consider the superpotential and expand the superpotential around the meson VEV as above. We ignore the constant term. For the potential expanded about the SUSY ground state, the term linear in $S$ must cancel, which yields the relation
\begin{equation}\label{Lam_mu_rel}
     \Lambda^{3N_c - N_f} = \mu^{N_c - N_f} v^{2N_c}
\end{equation}
For the remaining terms, with the required meson rescaling $\Mes \rightarrow \sqrt{2}\,v\Mes$, and by employing Eq.~(\ref{mes_def}) and $\Tr \Mes^2 = S^2 + \Tr M^2$, we arrive at
\begin{align}
\begin{split}
W &= \mu \left(\frac{N_c}{N_c-N_f} S^2+\Tr M^2\right)  \\ &\quad - \frac{\sqrt{2}\,\mu}{3v}\left(\frac{N_c(2N_c-N_f)}{\sqrt{N_f}(N_c-N_f)^2} S^3+3\frac{2N_c-N_f}{\sqrt{N_f}(N_c-N_f)} S \Tr M^2+2\Tr M^3\right) + \dotsb
\end{split}
\end{align}

Notice that the adjoint meson mass is twice the quark mass, $m\equiv2\mu$. Additionally, we can see the tree level mass ratio of the singlet to the non-singlet mesons
\begin{align}
r = \frac{N_c}{N_c-N_f} \,.\label{r}
\end{align}
We have also verified this result using an explicit parameterization of the quark superfields along the $D$-flat directions. 

Since we want the singlet to provide an s-channel resonance in adjoint meson scattering, we will require $r=2$ or $N_c = 2 N_f$. With the definition (\ref{r}) the superpotential becomes
\begin{align}
W = \frac{1}{2} m (r S^2 + \Tr M^2)  - \frac{\sqrt{2}\, \mu}{3v}\left( \frac{r(r+1)}{\sqrt{N_f}} S^3+3\frac{r+1}{\sqrt{N_f}} S \Tr M^2+2\Tr M^3 \right) + \dotsb 
\label{final_super}
\end{align}

In Appendix \ref{section:masscalc} we use Eq.~(\ref{kahler11}) and Eq.~(\ref{final_super}) to compute the one-loop mass corrections of the mesons. For the relevant case of $r=2$, the one-loop corrections give
\begin{align}
    r_\text{1-loop} = m_S/m \equiv 2(1-\delta) = 2  \left( 1 - \frac{\mu^2}{16 \pi^2 v^2} \frac{104 + 41 N_f^2}{N_f} \log \frac{v}{2\mu } \right) \label{eq:rLoop}
\end{align}
where $m_S$ is the mass of the singlet and we have introduced the mass splitting $\delta$. Note that we are being agnostic about the renormalization scheme as we are only interested in the log running. We see that the singlet sits below threshold so that the adjoint-adjoint scattering matrix element is maximal at zero velocity. Finally, we will refer to the dimensionless Yukawa coupling
\begin{align}
    y \equiv \mu /v.
\end{align}

While in this work we focus on the gauge group $SU(N_c)$, the phenomenology is qualitatively the same for the gauge groups $SO(N_c)$ with $N_f$ flavors and $Sp(2N_c)$ with $2N_f$ flavors. The different Dynkin indices change the power of $\det \Mes$ in the superpotential, and thus change Eq.~(\ref{r}). With $SO$ gauge groups one simply replaces $N_c \rightarrow N_c - 2$ while for $Sp$ the replacement is $N_c \rightarrow N_c + 1$.

\subsection{$U(1)_D$ gauge interaction}
\label{subsection:U(1)_D gauge interaction}

A dark photon, which we denote by $A'$, is the means by which the SM and dark sectors are coupled. It comes from gauging the $U(1)_D$ subgroup of the flavor symmetry group. There is then a kinetic mixing between the dark photon and the SM $U(1)_Y$ gauge boson. The SUSY breaking effects of this coupling are explored in the next subsection.

In our model we take $N_f$ to be even and a quark charge matrix $Q$ such that $Q^2 \propto \mathbf{1}$. This choice is sufficient to forbid a neutral meson from decaying to two dark photons (see Appendix \ref{anom_match} and \cite{Berlin:2018tvf}). Therefore we have an equal number of positively and negatively charged quarks. We also define $g_D$ as the $U(1)_D$ coupling strength, in a normalization where charged dark mesons have charge $\pm 1$.

The dark photon gives a one-loop mass renormalization to charged mesons. In the case of a dark photon with mass $m_{A'}$ below the cutoff $v$ this becomes
\begin{align}\label{SUSYsplit}
    m_\pm = m_0 \left(1 + \frac{g_D^2}{8\pi^2} \log \frac{v}{m_{A'}} \right)
\end{align}
In this paper we will generally take $g_D$ to be larger than the DM self-coupling $y$, which we will find in Section \ref{section:Self Interactions} to be $\sim 5\times 10^{-4}$. This mass correction is thus larger than the the mass shift in Eq.~(\ref{eq:rLoop}) and the charged dark matter is not expected to take part in the resonant interaction fundamental to our model. However, the analysis can be repeated when $g_D$ is small and the charged DM is approximately degenerate with the neutral DM. This case is qualitatively very similar. Therefore, the low velocity, resonant part of the DM self-scattering cross section is dominated by neutral DM, while the high velocity part will be dominated by dark photon exchange between charged DM (since the $U(1)_D$ coupling is greater than the DM self-coupling).

\subsection{Effect of SUSY breaking}
\label{subsection:Effect of SUSY breaking}

Supersymmetry, while critical for producing a resonance for neutral meson scattering, is not essential as an exact symmetry for a working model. Moreover, SUSY breaking is a necessary result of the coupling between the visible and hidden sectors. Since the visible sector is the MSSM with a sizeable amount of SUSY breaking, it is inevitable that at least some of this SUSY breaking will be transmitted to the hidden sector, even if we assume no direct coupling between the hidden sector and dynamical SUSY breaking sector.\footnote{In particular, we assume the gravity-mediation \cite{Alvarez-Gaume:1983drc,Hall:1983iz} or anomaly mediation \cite{Randall:1998uk,Giudice:1998xp} is negligible in the hidden sector. For example, low-energy gauge mediation \cite{Alvarez-Gaume:1981abe,Dine:1981gu} or vector mediation \cite{Hook:2015tra,Hook:2018sai} models belong to this category.} In this section we explain what the leading SUSY breaking effects in the hidden sector are, some of which will be important for the determination of the relic density and indirect detection constraints. 

The actual link to the visible sector will be as usual a small kinetic mixing between the dark photon and $U(1)_\text{Y}$ gauge boson $B$ of the (MS)SM. This kinetic mixing itself can be fully supersymmetric of the form  
\begin{align}
\epsilon \int d^{2}\theta W_{\text{D} \alpha} W_\text{Y}^\alpha + h.c. = \epsilon (- F_{\text{D} \alpha \beta} F_\text{Y}^{\alpha \beta} + 4 i \bar{\lambda}_\text{D}  \slashed{\partial} \lambda_\text{Y} + 2 D_\text{D} D_\text{Y})
\end{align}
where the $W^\alpha$s are field strength chiral superfields. Abelian vector superfields contain a photon, a photino, and an auxiliary $D$-term. Treating $\epsilon$ perturbatively, the largest effect on the dark spectrum comes from the mixing of the $D$-terms given by
\begin{align}
\mathcal{L} \supset \frac{1}{2} D_\text{D}^2 + D_\text{D} J_\text{D} + \frac{1}{2} D_\text{Y}^2 + D_\text{Y} J_\text{Y} + 2\epsilon D_\text{D} D_\text{Y}
\end{align}
where the $J$s are the usual $D$-terms $g \sum_i q_i |\phi_i|^2$ formed from the scalar components $\phi_i$ of the chiral superfields charged under the U(1)'s. After integrating out the auxiliary $D$-terms one obtains the Lagrangian for the physical degress of freedom. In the $\epsilon = 0$ case of no mixing one just obtains the usual two independent $D$-terms  (two terms quadratic in the respective $J$s), giving rise to the expected scalar quartic terms. However in the presence of mixing a third term is introduced, which at linear order in $\epsilon$ is equal to $2\epsilon J_\text{D} J_\text{Y}$. Importantly, $J_\text{Y}$ is quadratic in the MSSM Higgs VEV $v_h$ and generates a mass term for scalar fields charged under $U(1)_D$,
\begin{align}
2\epsilon \frac{g_\text{Y}}{2} (|H_u|^2 - |H_d|^2) g_\text{D} \sum_i q_i \phi_i^* \phi_i
\end{align}
where the $H$s are the up- and down-Higgses of the MSSM and the sum is over the dark degrees of freedom. Using the MSSM identities $|H_u|^2 - |H_d|^2 = \frac{v_h^2}{2} \cos 2\beta$ and $g_\text{Y} = \frac{e}{\cos \theta_\text{W}}$, this mass term is
\begin{align}
\approx \epsilon \frac{g_\text{D}}{e} \cos 2\beta (56 \text{GeV})^2 \sum_i q_i \phi_i^* \phi_i
\end{align}

Note that $\epsilon$, $g_\text{D}$, and $\beta$ are all free parameters. This shifts the mass-squared of charged bosons (no effect on fermions at tree level), so that combined with the result of the previous subsection we have
\begin{align}
    m_{\pm,\text{bos}}^2 = m_0^2 \left(1 + \frac{g_D^2}{4\pi^2} \log \frac{v}{m_{A'}} \pm \epsilon \frac{g_\text{D}}{e} \cos 2\beta \frac{(56 \text{GeV})^2}{m_0^2} \right)
    \label{eq:SUSY_Mass_Splitting}
\end{align}

The competition between the SUSY preserving and SUSY breaking term leads to two situations. When the SUSY preserving term dominates, we have lighter neutral DM and heavier charged DM. When the SUSY breaking term dominates the lightest particles in the dark sector are charged bosons, and $U(1)_D$ charge conservation ensures their stability. It is important in this case for the neutral DM not to be too much heavier, otherwise it will be depleted and unable to participate in the resonant self-interaction. For simplicity we will therefore insist that the lightest dark particle (LDP) is neutral, whereby the SUSY preserving term must be larger than the SUSY breaking term\footnote{This is a somewhat conservative bound. It is possible to be above this bound if the dark QED and SUSY breaking contributions to the charged bosons mass cancel. This would allow the lightest charged bosons to be the SIDM as it would be resonantly enhanced. We don't discuss this situation further.}. In fact, we are interested in $\epsilon \approx 10^{-12}$ as we will discuss later (see Fig.~\ref{fig:Essig_keV_Results}), and hence this is indeed the case.

\begin{figure}[t]
    \centering
    \begin{tikzpicture}
    \begin{feynman}
    \vertex  (i) at (0,0);
    \vertex  (f) at (9,0);
    \vertex  (a) at (3,0);
    \vertex  (b) at (6,0);
    \vertex  (c) at (5.45,1.06);
    \vertex  (d) at (3.55,1.06);

    \diagram*{
    (i)  --[charged scalar, edge label' = $M$] (a) --[fermion, edge label' = $\psi_M$]  (b) --[charged scalar, edge label' = $M$] (f) ,
    (b) --[photon,fermion, out=90,in = -45,edge label' = $\lambda_D$] (c),
    (c) --[photon,fermion,out = 135, in = 45,edge label' = $\lambda_B$] (d),
    (d) --[photon,fermion, out = 225, in = 90,edge label' = $\lambda_D$] (a),
    };
    \node at (5.45,1.06) {\Large$\otimes$};
    \node at (3.55,1.06) {\Large$\otimes$};
    \end{feynman}
    \end{tikzpicture}
    \caption{SUSY breaking contribution from dark charged meson mass correction.}
    \label{fig:Charged_Meson_Mass_Correction}
\end{figure}
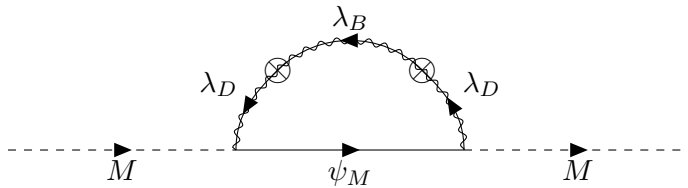

All other effects of the $D$-term on the dark spectrum are subleading. The largest additional correction is expected to come from the large SUSY breaking bino mass $m_{\tilde{B}}$. Adding a (SUSY preserving) St\"{u}ckelberg mass $m_D$ to the dark photon we obtain the gaugino mass  matrix 
\begin{align}
\begin{pmatrix}
\slashed{p} - m_D & 2\epsilon \slashed{p} \\
2\epsilon \slashed{p} & \slashed{p} - m_{\tilde{B}}
\end{pmatrix}
\end{align}
This yields a corrected dark photino mass $m_D (1 - 4\epsilon^2 \frac{m_D}{m_{\tilde{B}}-m_D})$ at leading order in $\epsilon$. Due to the $\epsilon^2$ suppression of the correction and the expected smallness of the dark photon mass in comparison to the SM photino mass, the effect on the dark meson spectrum is negligible. A more important source of SUSY breaking is a dark charged meson mass correction from the diagram in Figure \ref{fig:Charged_Meson_Mass_Correction},
\begin{align}
\Delta m^2 \propto \frac{\alpha_\text{D}}{4 \pi} \epsilon^2 m_{\tilde{B}}^2 \log \frac{m_{\tilde{B}}^2}{m^2}
\end{align}
As a loop correction at order $\epsilon^2$, this is negligible when compared with the tree level $D$-term SUSY breaking. 

\section{Self-Interactions}
\label{section:Self Interactions}

\begin{figure}[t]
    \centering
    \includegraphics[scale = 0.5]{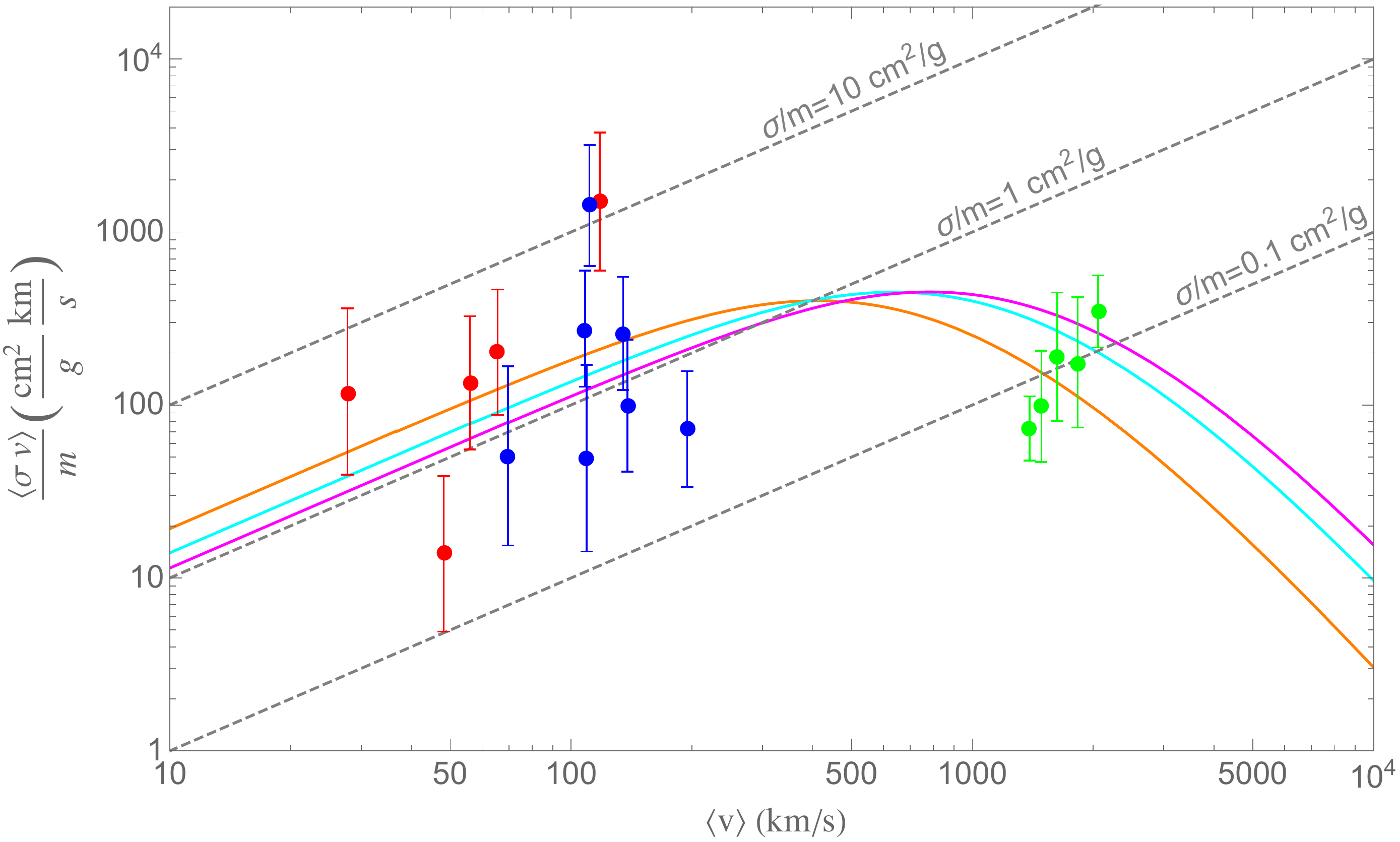}
    \caption{The velocity dependence of the DM self-interaction cross section. The red, blue, and green data points come, respectively, from the analysis of dwarf galaxies, LSB spiral galaxies, and galaxy clusters. These are taken directly from \cite{Kaplinghat:2015aga}. There also exist error bars for $\expval{v}$, which are relevant for the $\chi^2$ test in Figure \ref{fig:ParameterSpace}, but these are not shown. The dashed gray lines represent the values of $\expval{\sigma v}/m$ if $\sigma/m$ were the velocity independent values labeled by the line. The orange, cyan, and magenta curves correspond to the points in parameter space shown in Figure \ref{fig:ParameterSpace}.}
    \label{fig:DM_Halos}  
\end{figure}

As mentioned in the introduction, one of the key features of SRDM is that it allows for velocity dependent self-interactions capable of resolving small scale puzzles seen in the dark matter halos of dwarf galaxies. DM halos seen in dwarf galaxies, low surface brightness (LSB) spiral galaxies, and relaxed galaxy clusters whose halos are appropriately spherical  were studied in \cite{Kaplinghat:2015aga}. There, specific ranges of the cross section at given average relative velocities were shown to be a potential resolution to the small scale puzzles. This data is included in Figure \ref{fig:DM_Halos}. In this section, we calculate the resonant self-interaction cross section and find the parameter space of the model which is consistent with their analysis.

In our model, we will have two general types of dark matter: those charged under $U(1)_D$ and those that are neutral. The charged DM, as we shall see, will be completely negligible when it comes to self-interactions in dark matter halos, as their dark QED mass corrections of order $\alpha_D$ prevent a maximally enhanced resonant cross section. We shall also see that the abundance of charged DM is suppressed relative to the neutral DM because of the dark QED mass correction.

To compare the effectiveness of the SRDM model in describing small scale puzzles in dark matter halos, we must calculate the thermally averaged cross section $\expval{\sigma v}$, where the incoming and outgoing particles are neutral DM. Using only neutral relic DM is an assumption that will be justified a posteriori for a wide range of dark photon couplings. However, for large values of $\alpha_D$ we will find that a sizeable fraction of the DM is charged. Including these extra degrees of freedom and their dark QED interactions will lead to at most $\mathcal{O}(1)$ changes in the best fit values of $m$ and $y$.

For all of the experimental data of interest, we are well within the non-relativistic regime. We also assume the self-interactions are strong enough such that the velocity distribution is roughly Boltzmannian. Then the thermally averaged cross section can be written as
\begin{align}{}
    \expval{\sigma v} \approx  \frac{1}{(\sqrt{2\pi}v_0)^3}\int d^3 v \sigma_\text{int}(v) v e^{-v^2/2v_0^2} \label{sigma}
\end{align}
where $v$ is the relative speed of the incoming particles and the mean relative speed is given by $\expval{v}=\sqrt{8}v_0/\sqrt{\pi}$. We have also defined $\sigma_\text{int}$ to be the average two-to-two cross section given by 
\begin{align}
    \sigma_\text{int}(v) \approx \frac{1}{64 \pi m^2}\frac{1}{2! \mathcal{N}_\text{neutral}^2}\sum_{i_1,i_2,f_1,f_2}^{\text{neutral}}|\mathcal{M}_{i_1,i_2\rightarrow f_1 f_2}|^2
\end{align}
where the indices $i_1,i_2,f_1,f_2$ go over all neutral DM states (including helicities), and $\mathcal{N}_\text{neutral}$ is the total number of such states. We have also made the approximations $s\rightarrow 4m^2$, $t\rightarrow 0$, $u \rightarrow 0$; however, we must be careful to preserve the terms in the denominator of the form $(s-m_S^2)$. We can therefore write the cross section as
\begin{align}\label{crosssec_form}
    \sigma_\text{int}(s) = \sigma_0+\frac{\sigma_1 m^2}{s-m_S^2} +\frac{\sigma_2 m^4}{(s-m_S^2)^2} 
\end{align}
In the denominator, we must be a little more careful with our non-relativistic approximation and write $s \approx 4m^2 + m^2 v^2$. Using $m_S= 2m(1-\delta)$ we get $s - m_S^2 \approx m^2(8\delta + v^2)$. Since both $\delta $ and $v^2$ will be very small in the non-relativistic regime, the final term of Eq.~(\ref{crosssec_form}) is dominant.

The thermally averaged cross section is given by
\begin{align}
    \expval{\sigma v} \approx&
    \frac{16\sigma_2}{\pi^2\expval{v}^{3}} \times \left(-1-(1+\xi)e^{\xi}\text{Ei}(-\xi)\right) \quad \text{with} \quad
    \xi = \frac{32\delta}{\pi \expval{v}^2}
\end{align}
where $\text{Ei}(-\xi)$ is the exponential integral function. For the purpose of intuition, it is helpful to note the following limiting behavior
\begin{align}
    \expval{\sigma v} \approx&
    \begin{cases}
    \frac{\sigma_2 \expval{v}}{64 \delta^2} &  \expval{v}\ll \sqrt{\delta} \\
    \frac{16\sigma_2}{\pi^2\expval{v}^{3}}\left(\log\left(\frac{\pi \expval{v}^2}{32\delta}\right) - \gamma_M - 1\right) &  \sqrt{\delta}\ll \expval{v} \ll 1 
    \end{cases}
\end{align}
where $\gamma_M$ is the Euler-Mascheroni number. Using the data in Figure \ref{fig:DM_Halos} to estimate the low- to high-velocity transition to be around $v_\text{trans} \sim 500 \text{ km/s}$, we can estimate that $\delta = v_\text{trans}^2/8 \sim 10^{-7}$. For $N_f = 2$, this will imply $y \sim 10^{-3}$.

Also, since $\sigma_2 \propto y^4/m^2$ and $\delta  \propto y^2\log(y^{-1})$, we see that for fixed $N_f$ the thermally averaged cross section is a function of $\expval{v}$ with only two parameters, $m$ and $y$. This is different than many other SIDM models where the mass splitting depends on an additional continuous parameter. 
Using only neutral DM external particles, $\sigma_2$ can be calculated by hand or with a combination of FeynRules \cite{FeynRules1,FeynRules2}, FeynArts \cite{FeynArts}, and FormCalc \cite{FormCalc}. The result is
\begin{align}
    \sigma_2 = \frac{8}{\pi  N_f^2}\frac{y^4}{m^2}. \label{eq:sigma2}
\end{align}

\begin{figure}[t]
        \centering
        \begin{subfigure}[b]{0.32\textwidth}
            \centering
           \includegraphics[width=\textwidth]{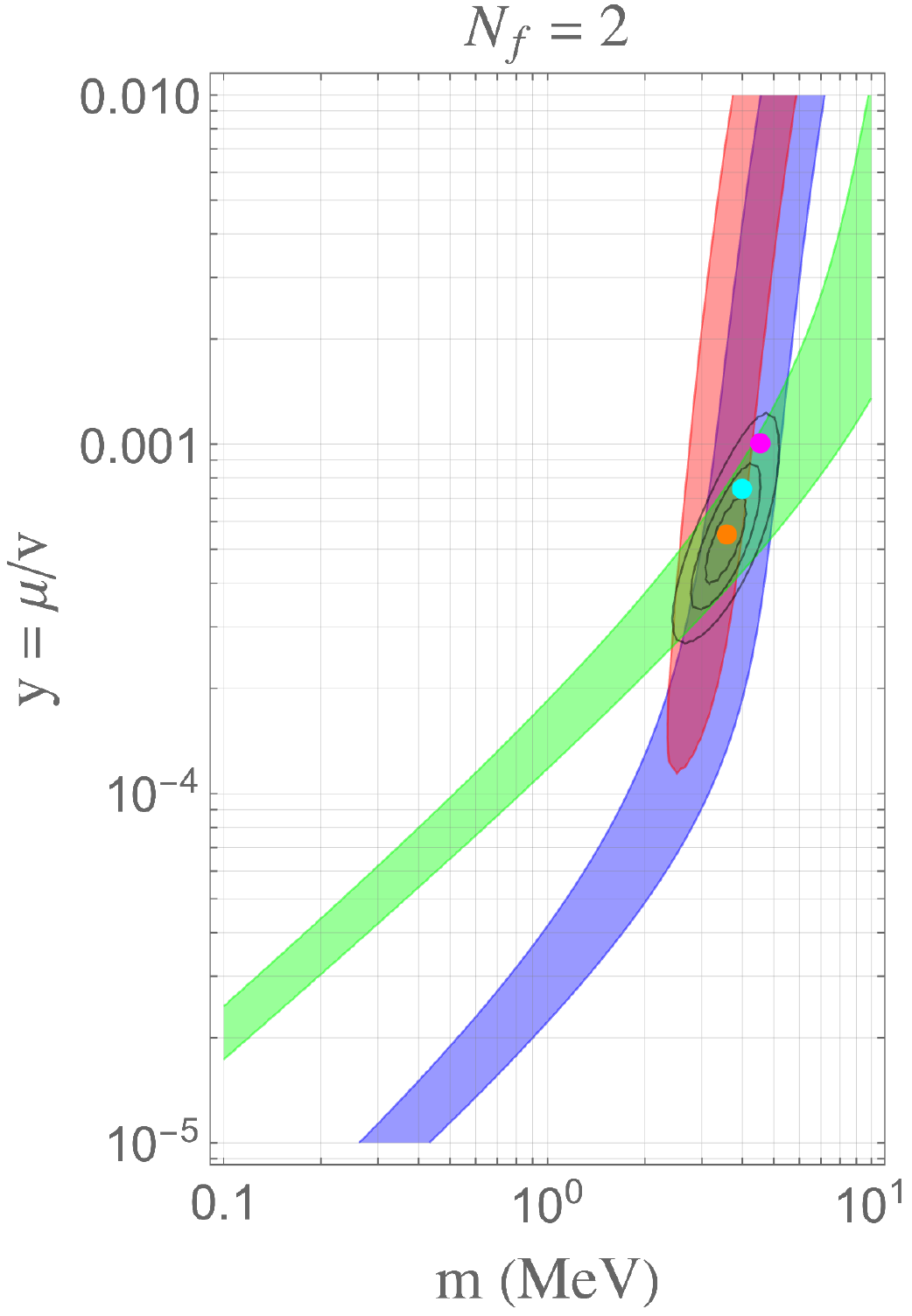} 
        \end{subfigure}
        \hfill
        \begin{subfigure}[b]{0.32\textwidth}   
            \centering 
            \includegraphics[width=\textwidth]{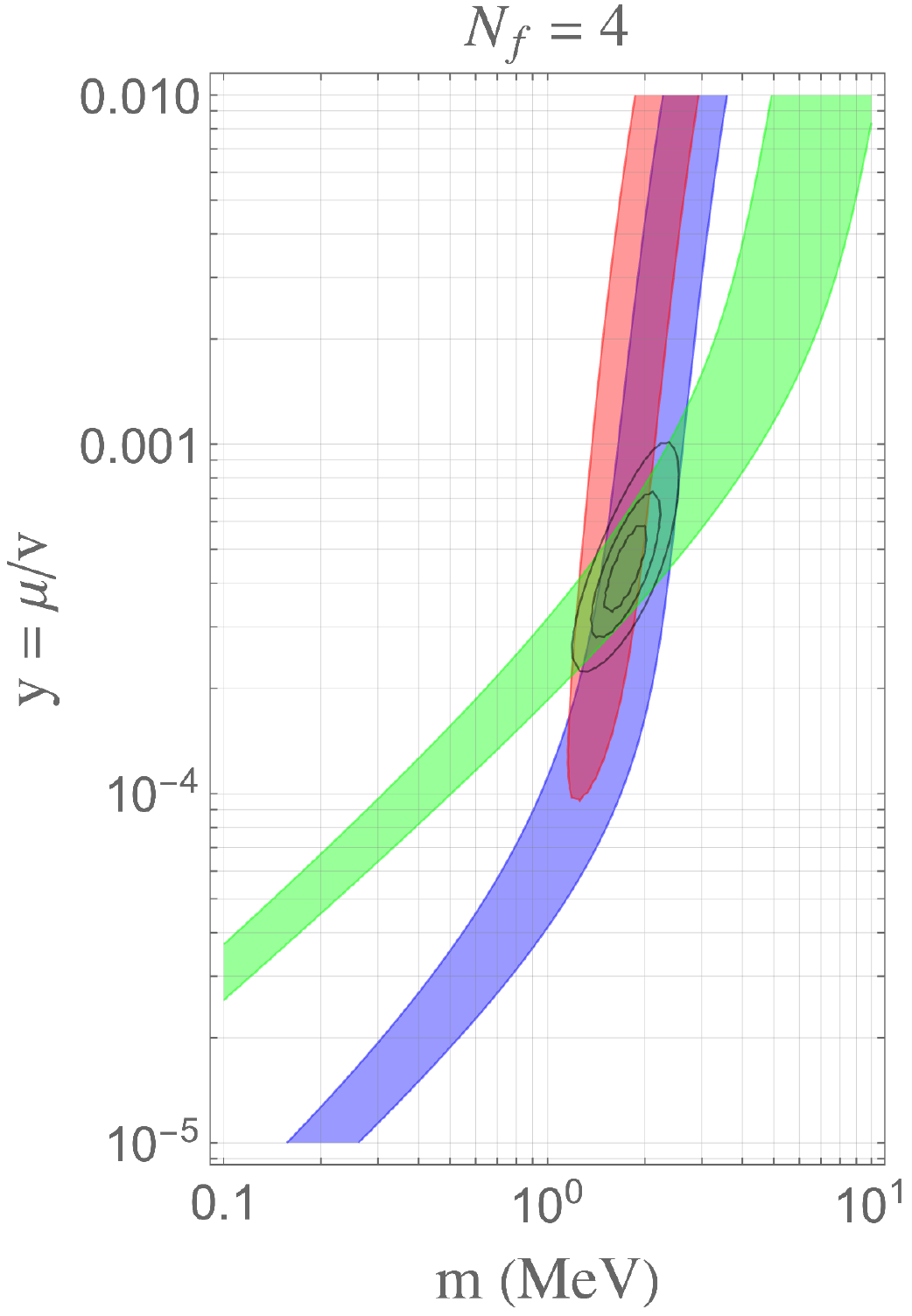} 
        \end{subfigure}
        \hfill
        \begin{subfigure}[b]{0.32\textwidth}   
            \centering 
            \includegraphics[width=\textwidth]{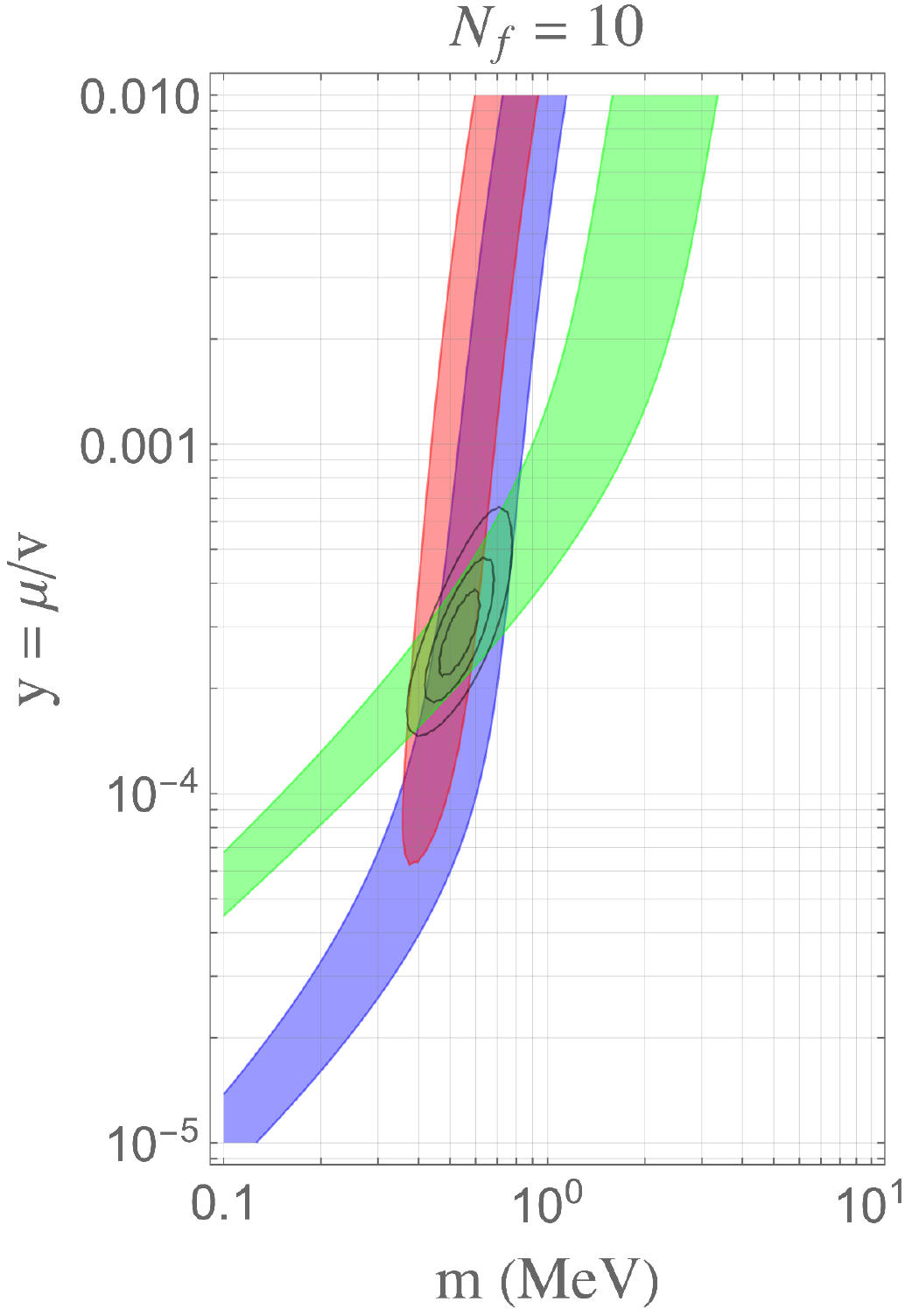} 
        \end{subfigure} 
        \caption{Results of a $\chi^2$-fit to the data from \cite{Kaplinghat:2015aga} for Super-Resonant Dark Matter with $N_f = 2$, $N_f = 4$ and $N_f = 10$. The red, blue, and green shaded regions are the 95\% confidence level regions for fitting to the data from dwarf galaxies, LSB spiral galaxies, and galaxy clusters, respectively. The black contours correspond to 90\%, 95\%, and 99\% confidence levels for the equally weighted combination of all three data sets. The orange, cyan, and magenta dots on the $N_f = 2$ plot correspond to the curves shown in Figure \ref{fig:DM_Halos}. We pick some of these slightly offset from the best fit point to show variation in Figure \ref{fig:DM_Halos}. The actual best fit points are $(m =3.55 \text{ MeV} ,y=5.2\times 10^{-4})$ for $N_f = 2$,$(m =1.74 \text{ MeV},y=4.4\times 10^{-4})$ for $N_f = 4$, and $(m = 0.54\text{ MeV},y=2.8\times 10^{-4})$ for $N_f = 10$.}
        \label{fig:ParameterSpace}
    \end{figure}

The $\chi^2$-fits of the thermally averaged cross section for $N_f=2,4,10$ to the data are shown in Figure \ref{fig:DM_Halos}. The viable parameter space around the best fit values are shown in Figure \ref{fig:ParameterSpace}. The main result is that the dark matter has a mass in the MeV range and is weakly coupled, as expected. This mass range and coupling are also quite constrained.

For example, for $N_f = 2$ we find best fit values of $m =3.55 \text{ MeV}$ and $y=5.2\times 10^{-4}$. However, a reasonable range for these parameters is a mass between $2.5 - 5 $ MeV and a self-interaction coupling between $3\times 10^{-4} - 10^{-3}$. The relation $\Lambda = y^{-4/5}\mu$ (see Eq.~(\ref{Lam_mu_rel})) tells us the dynamical scale of the gauge theory sits around a few GeV. Taking larger values of $N_f$ while maintaining the mass splitting $\delta$ fixed, Eq.~(\ref{eq:rLoop}) tells us that $y$ must scale as $N_f^{-1/2}$. To maintain $\expval{\sigma v}/m$ the mass must therefore scale as $N_f^{-4/3}$. Thus, we cannot take arbitrarily high values of $N_f$ due to a lower bound on the mass from the neutral LDP condition (below).

\section{Relic Abundance}
\label{section:Relic Abundance}

We have so far presented a model with a natural resonance that is capable of explaining the small scale puzzles. The final step in producing a viable dark matter model is explaining the present-day relic abundance. Beyond this, it must be ensured that an $\mathcal{O}(1)$ fraction of the relic DM is neutral and can participate in the resonant interaction (for small enough $\alpha_D$ the charged DM can participate too).

In this section we constrain the parameters of the model by trying to produce the correct relic abundance for both freeze-out and freeze-in scenarios, which we describe below (detailed calculations can be found, for example, in \cite{Essig_keV}). We find that freeze-out is excluded, while freeze-in is viable.

In both scenarios, a dark photon lighter than the dark mesons is ruled out as the mesons would down-scatter to dark photons, which do not experience the desired resonant self-interaction. For simplicity we will further assume throughout that $m_{A'}>2m$ so that an on-shell dark photon can decay to two dark mesons.

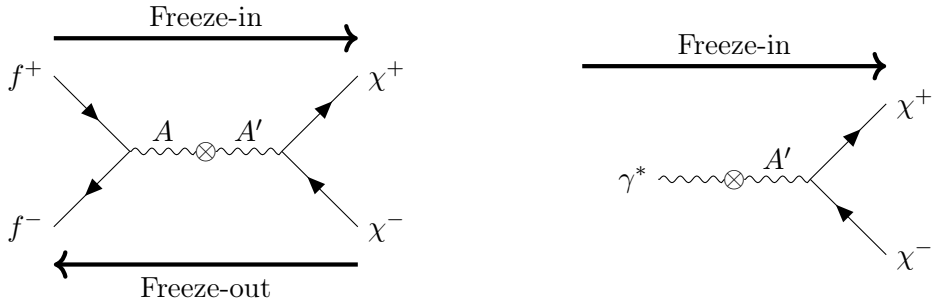
\begin{figure}[h]
    \centering
    \begin{subfigure}{.475\textwidth}
    \centering
    \begin{tikzpicture}
    \begin{feynman}
    \vertex [label=left:\(f^+\)] (i1) at (0,1);
    \vertex [label=left:\(f^-\)] (i2) at (0,-1);
    \vertex  (a) at (1,0);
    \vertex  (b1) at (1.87,0);
    \vertex  (b2) at (2.13,0);
    \vertex  (c) at (3,0);
    \vertex [label=right:\(\chi^+\)] (f1) at (4,1);
    \vertex [label=right:\(\chi^-\)] (f2) at (4,-1);
    \node at (2, 0)   (otimes) {$\otimes$};
    
    \node at (2, 1.8)   (otimes) {Freeze-in};
    
    \draw[ultra thick,->] (0,1.5)-- (4,1.5);
    
    \node at (2, -1.8)   (otimes) {Freeze-out};
    \draw[ultra thick,->] (4,-1.5)-- (0,-1.5);

    \diagram*{
    (i1)  --[fermion] (a) --[fermion]  (i2) ,
    (a) --[photon,edge label = $A$] (b1),
    (b2)--[photon,edge label = $A'$] (c),
    (f1) --[anti fermion] (c) --[anti fermion] (f2) 
    };
    \end{feynman}
    \end{tikzpicture}
\end{subfigure}
\begin{subfigure}{.475\textwidth}
    \centering
    \begin{tikzpicture}
    \begin{feynman}
    \vertex  [label=left:\(\gamma^*\)] (a) at (1,0);
    \vertex  (b1) at (1.87,0);
    \vertex  (b2) at (2.13,0);
    \vertex  (c) at (3,0);
    \vertex [label=right:\(\chi^+\)] (f1) at (4,1);
    \vertex [label=right:\(\chi^-\)] (f2) at (4,-1);
    \node at (2, 0)   (otimes) {$\otimes$};
    \node at (2, 1.8)   (otimes) {Freeze-in};
    
    \draw[ultra thick,->] (0,1.5)-- (4,1.5);
    \diagram*{
    (a) --[photon] (b1),
    (b2)--[photon,edge label = $A'$] (c),
    (f1) --[anti fermion] (c) --[anti fermion] (f2) 
    };
    \end{feynman}
    \end{tikzpicture}
\end{subfigure}
\caption{Diagrams for freeze-out and freeze-in via a dark photon.}
    \label{fig:Plasmon_Decay}
\end{figure}

The first mechanism we consider to produce the DM relic abundance is freeze-out. In this scenario the SM and dark sectors of the early universe are in thermal equilibrium. The dominant equilibrating process is pictured on the left of Figure \ref{fig:Plasmon_Decay}, where $\chi^\pm$ denotes the charged fermionic mesons and $f^\pm$ SM fermions (electrons). The contributions from bosonic mesons are subdominant because their cross section is p-wave suppressed, having an additional factor of $v^2$.\footnote{One subtlety: For the range of $\epsilon$ relevant to freeze-out, the SUSY breaking mass correction of Eq.~\eqref{eq:SUSY_Mass_Splitting} dominates SUSY preserving term. Therefore, the density of the lighter charged bosonic DM is Boltzmann enhanced when compared with the heavier fermions. This can overcome the p-wave suppression and thus put charged bosons in control of the freeze-out process. However, they will be also be Boltzmann enhanced with respect to the neutral DM and the small relative fraction of neutral DM in the dark sector will be unable to reproduce the small scale anomalies.} Note the factor of $\epsilon^2$ arising from dark-SM photon mixing. As the universe cools below the DM mass, its number density becomes Boltzmann suppressed. This suppression eventually lowers the rate of the equilibrium process below the Hubble rate, at which time we say that the dark sector freezes out from the SM sector. With no more DM number changing processes, its abundance is fixed.

\begin{figure}[t!]
    \centering
    \includegraphics[scale = 0.5]{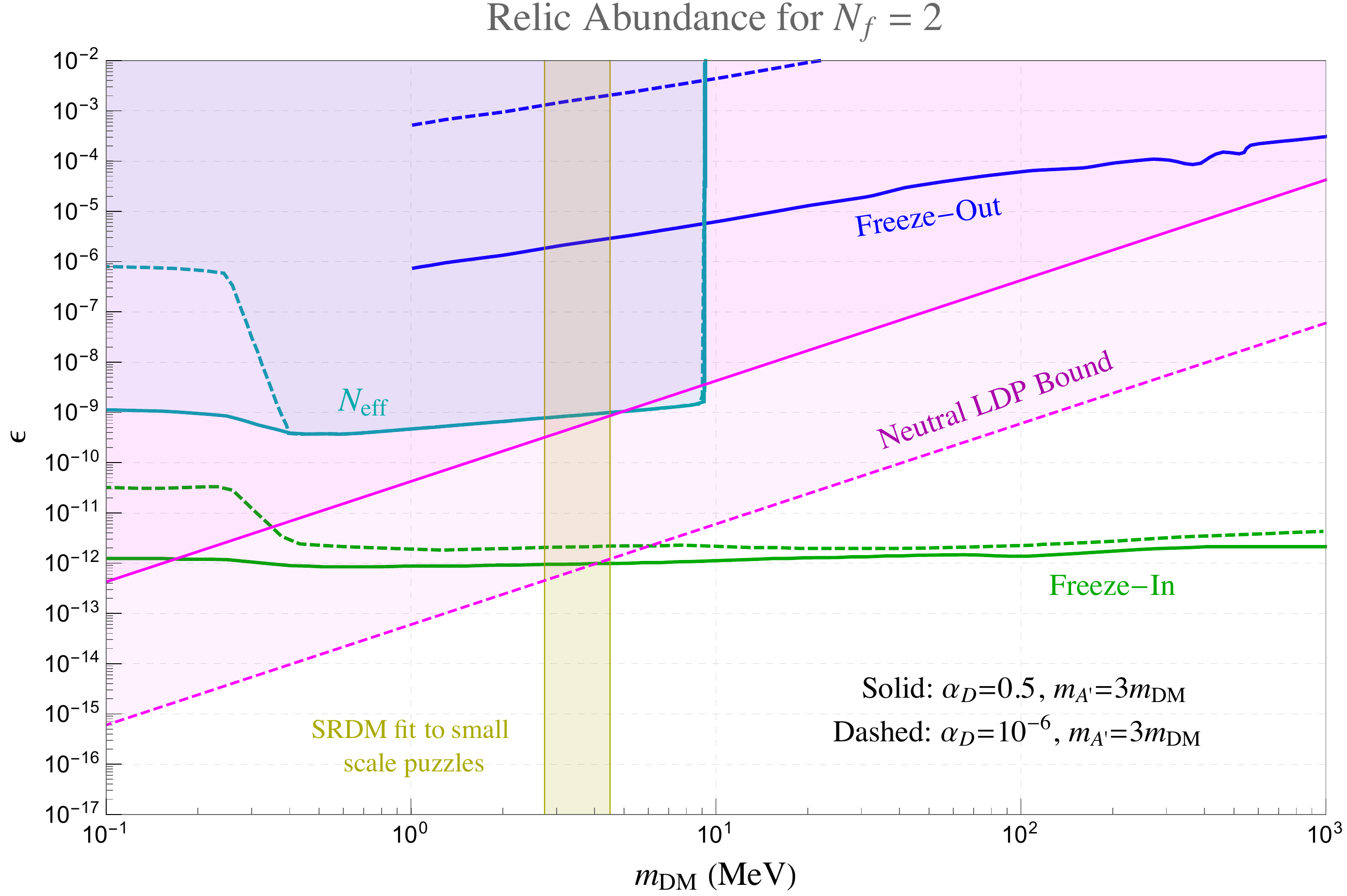}
    \caption{Constraints on kinetic mixing $\epsilon$ as a function of dark matter mass for $N_f=2$ and $m_{A'}=3m$. The blue freeze-out scenario relic abundance line, green freeze-in relic abundance line, and teal $N_\text{eff}$ bound (constraint from effective number of degrees of freedom present during BBN) are taken from \cite{Essig_keV}. In magenta, we have added the neutral LDP bound requiring the neutral DM is lighter than the charged DM. Note that we have assumed the largest possible SUSY breaking correction with $|\cos 2\beta| = 1$, which arises in the large $\tan \beta$ limit.  The vertical gold-shaded area indicates the region of parameter space where the SRDM model best fits the data from \cite{Kaplinghat:2015aga}. The plot exhibits very weak $N_f$ dependence; in particular, the neutral LDP bound and freeze-in lines do not depend on $N_f$ while the DM mass decreases for increasing $N_f$.}
    \label{fig:Essig_keV_Results}
\end{figure}

The parameters leading to the correct freeze-out relic abundance for $N_f=2$ and $m_{A'} = 3m$ are plotted in blue in Figure \ref{fig:Essig_keV_Results}, which has been adapted from \cite{Essig_keV}. In the mass range found in Section \ref{section:Self Interactions} (vertical gold band), and for $\alpha_D = 0.5$, one finds that $\epsilon \sim 10^{-6}$. A smaller dark photon coupling necessitates a larger $\epsilon$. Also shown is the teal-colored $N_\text{eff}$ bound on the effective number of degrees of freedom during Big Bang Nucleosynthesis (BBN). This bound lies well below the freeze-out line in the mass range of interest and therefore eliminates the possibility of a freeze-out scenario in our model. The freeze-out scenario is also excluded by the aforementioned neutral LDP condition (magenta). Choosing larger $N_f$ or $m_{A'}$ does not remedy the situation.

The second possibility we explore is the freeze-in scenario. In this case the dark and SM sectors are never in equilibrium, and the dark sector is populated by decays from the SM. Specifically, SM particles annihilate to produce dark meson pairs as pictured on the left of Figure \ref{fig:Plasmon_Decay}. This process is dominated by the contribution at temperatures of order $m_{A'}$, when the dark photon is on-shell. One can think of this as two-processes in which a dark photon is produced (suppressed by $\epsilon^2$) and then decays overwhelmingly to dark mesons. Plasmon decay, shown on the right of Figure \ref{fig:Plasmon_Decay}, also contributes to freeze-in \cite{Dvorkin:2019zdi} but is only important at lower DM masses.

\begin{figure}[t!]
\centering
  \includegraphics[width=0.49\textwidth]{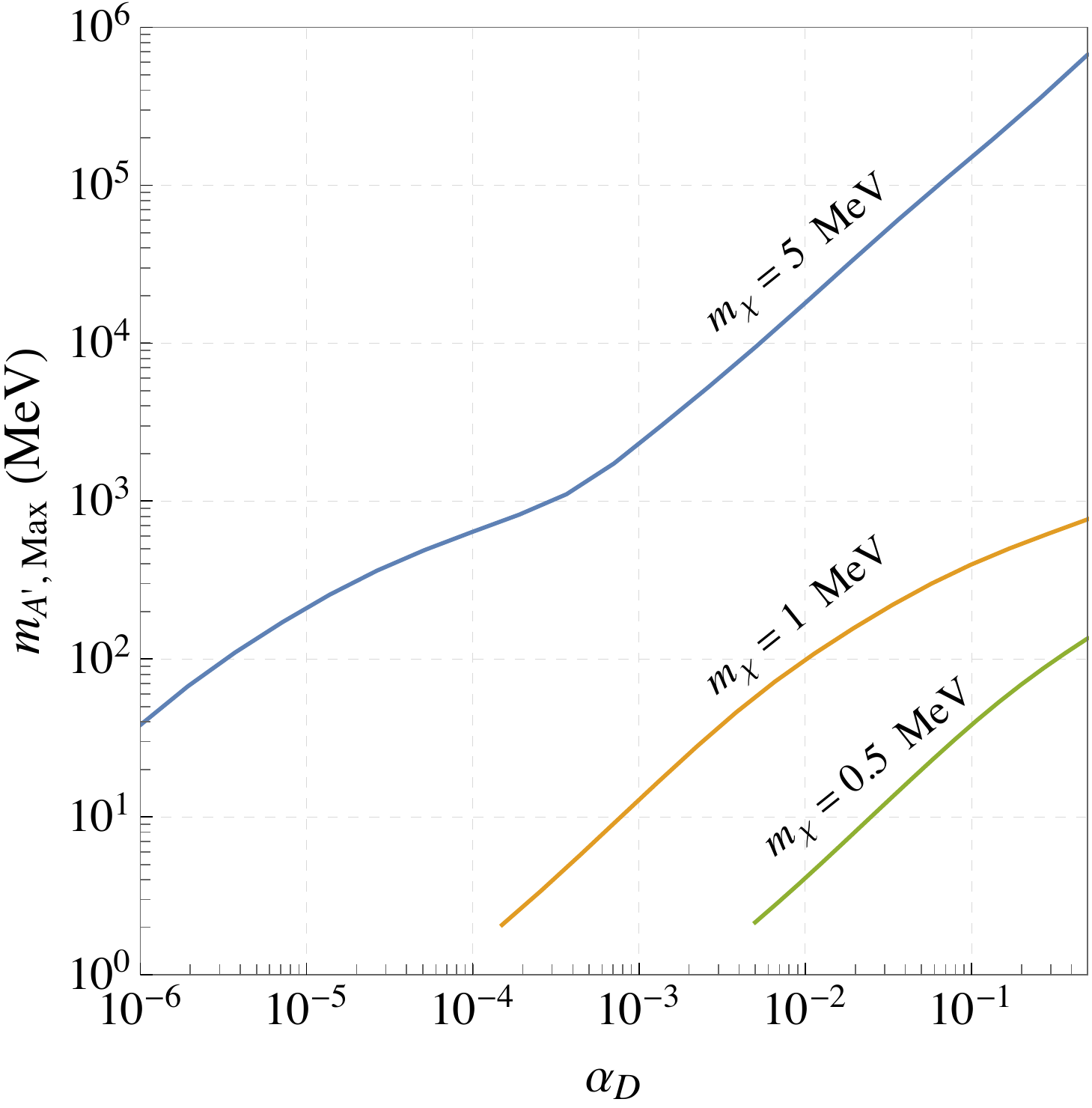}
  \includegraphics[width=0.49\textwidth]{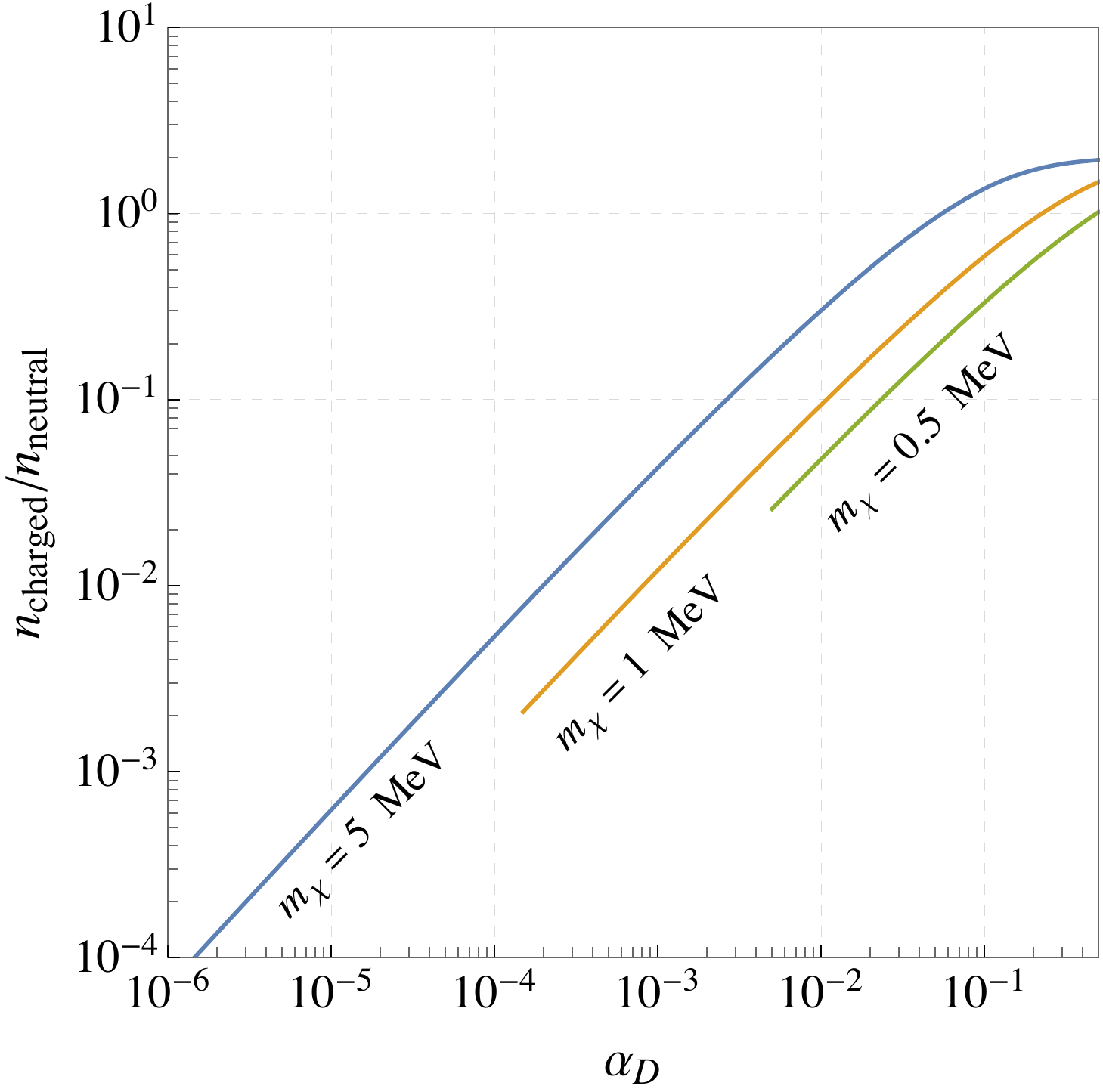}
    \caption{{\it Left:} Maximum value of the dark photon mass as a function of the dark photon coupling $\alpha_D$, for several different values of the dark meson mass. Lower meson masses are applicable for a great number of flavors $N_f$. {\it Right:} 
  The ratio of charged to neutral DM density for $N_f = 2$ after mutual decoupling, as a function of $\alpha_D$. A smaller coupling leads to a smaller charged mass splitting and therefore a more resonant self-interaction. As a result, a greater amount of Boltzmann suppression is required before the DM decouples.}
    \label{fig:MA_Max}

\end{figure}

The parameters producing the correct relic abundance in the freeze-in scenario were calculated\footnote{Note that we use the authors' calculation of dark photon decay to fermions, while in our model there are dark fermions and bosons. However because freeze-in is dominated by on-shell dark photons (that overwhelmingly decay into the dark sector), the difference is unimportant on the scale of Figure \ref{fig:Essig_keV_Results}.} in \cite{Essig_keV} and are also plotted in green in Figure \ref{fig:Essig_keV_Results}. The result is a viable dark matter model with $\epsilon \sim 10^{-12} \times (m_{A'}/3m_\chi)^{1/2}$. While only $m_{A'} = 3m$ is plotted, there is in fact a wide range of viable dark photon masses. The upper bound on the dark photon mass comes from the fact that both the freeze-in and $N_\text{eff}$ lines in the mass range of interest move up with dark photon mass as $\epsilon \propto m_{A'}^{1/2}$. However, the neutral LDP bound exhibits only weak dependence on $m_{A'}$. For $\alpha_D = 0.5$ we have $m_{A'} \lesssim \text{TeV}$ while for smaller $\alpha_D$ the bound decreases. For a number of dark meson masses, the maximum viable dark photon mass as a function of $\alpha_D$ is shown in the left panel of Figure \ref{fig:MA_Max}.

There is one important subtlety to take care of in the freeze-in scenario. The dark photon decays only to charged DM, which is heavier than the neutral DM and does not participate in the resonance. We must therefore ensure that the dark sector is at one point in equilibrium with itself so that the charged DM can partially down-scatter to neutral DM. How much charged DM remains will depend on when the dark sector leaves equilibrium with itself. We make these estimates below.

Assuming a dark photon of mass $m_{A'} > 2 m$, the dark matter is produced with temperature\footnote{Properly we should say energy and not temperature. However, we check that the dark sector comes into equilibrium with itself and so use the terms interchangeably.} $T_{\rm DM} \sim T \sim m_{A'}/2$, where $T$ is the SM sector temperature. Redshifting in a radiation dominated era means that $p \propto T$, so that while the DM remains relativistic we have $T_{\rm DM} \sim T$. Once the temperature crosses the DM mass threshold, $v \approx p/m \sim T/m$. Therefore $T_{\rm DM} \sim m v^2 \sim T^2 / m$. From Eq.~\eqref{crosssec_form} we have $\sigma \approx \sigma_2 / v^4$ (before the charged meson mass splitting becomes relevant---what we call `saturation of the resonance') and thus
\begin{equation}
    \expval{\sigma v} \propto \frac{\sigma_2}{\expval{v}^3} \propto T^{-3}
\end{equation}
On the other hand, the dark matter number density at temperature $T$ after the dark matter has frozen in goes like
\begin{equation}
    n(T) = \frac{\Omega_{{\rm DM},0}\, \rho_{c,0}}{m} \frac{s(T)}{s_0} \propto T^3
\end{equation}
where $\Omega_{{\rm DM},0}$ is the present-day energy fraction of dark matter, $\rho_{c,0}$ is the critical energy density, and $s$ is entropy density. Therefore, until the resonance saturates, the scattering rate in the dark sector $\Gamma = n \expval{\sigma v}$ is temperature independent. We have checked that for all $\alpha_D \lesssim 0.5$ this exceeds the Hubble rate, which decreases like $T^2$, at some temperature before the resonance saturates. The neutral and charged DM do indeed equilibrate.

The resonance saturates once $m v^2\sim 8 (m_\pm - m_0) $ (recall the argument of Section \ref{section:Self Interactions}) at which point the cross section no longer increases as the velocity decreases. At this time the Boltzmann suppression of the charged DM abundance relative to neutral DM abundance begins. Therefore, the charged to neutral annihilation rate will decrease exponentially with decreasing temperature. Once $\Gamma$ falls below the Hubble rate at dark matter temperature $T_{D,\text{freezeout}}$, the charged DM is frozen at relative abundance $\exp(-(m_\pm - m_0) / T_{D,\text{freezeout}})$, shown in the right panel of Figure \ref{fig:MA_Max}. For $\alpha_D \lesssim 0.01$ the charged DM forms a small fraction of the total relic DM. This justifies a posteriori not including the charged DM interactions in the self-interaction calculation for a wide range of dark photon couplings.

\section{Conclusion}
\label{section:Conclusion}

We have proposed a SIDM model based upon supersymmetric QCD. Due to SUSY and flavor symmetry, the low energy effective theory exhibits a natural 2:1 mass ratio leading to a velocity dependent self-interaction cross section. Fitting to self-interaction constraints and small scale puzzles fixes a weak self-coupling and a DM mass in the few MeV range. We find that while freeze-out is excluded, the freeze-in scenario can reproduce the DM relic abundance. For a dark photon with a mass just above the DM mass, we find a kinetic mixing of $\epsilon \sim 10^{-12}$, while with a heavier dark photon $\epsilon$ can be a few orders of magnitude larger. Finally, $\alpha_D$ controls the fraction of charged DM, which is $\mathcal{O}(1)$ for large $\alpha_D$ and exponentially suppressed for small $\alpha_D$.

\begin{acknowledgments}

AG and KL thank Simon Knapen for useful discussions. CC and AG are supported in part by the NSF grant PHY-2014071. CC, AG and EK are supported in part by the US-Israeli Binational Science Foundation grant 2020220. AG is also supported in part by the NSERC PGS-D fellowship. YH is supported in part by the Israel Science Foundation (grant No. 1112/17),  by the Azrieli Foundation and by an ERC STG grant (grant No. 101040019). YH and HM are also supported in part by the US-Israeli Binational Science Foundation (grant No. 2018140). EK is supported by the US-Israeli Binational Science Foundation (grants 2016153 and 2020220), the Israel Science Foundation (grant No. 1111/17), and by the I-CORE Program of the Planning and Budgeting Committee (grant No. 1937/12). HM is supported in part by the Director, Office of Science, Office of High Energy Physics of the U.S. Department of Energy under the Contract No. DE-AC02-05CH11231, by the NSF grant PHY-1915314, by the JSPS Grant-in-Aid for Scientific Research JP20K03942, MEXT Grant-in-Aid for Transformative Research Areas (A) JP20H05850, JP20A203, by WPI, MEXT, Japan, the Institute for AI and Beyond of the University of Tokyo, and Hamamatsu Photonics.  This project has received funding from the European Research Council (ERC) under the European Union’s Horizon Europe research and innovation programme (grant agreement No. 101040019). Funded by the European Union. Views and opinions expressed are however those of the author(s) only
and do not necessarily reflect those of the European Union. The European Union cannot be held responsible for them.

\end{acknowledgments}

\appendix
\section*{Appendix}

\section{1-Loop Mass Ratio Calculation}\label{section:masscalc}

In this Appendix we present the details of the one loop adjoint and singlet meson mass corrections. Recall that the UV $SU(N_c)$ gauge theory is broken to $SU(N_c-N_f)$ by the squark VEVs $v$. At this scale the gauge bosons corresponding to the broken generators eat some of the quark superfields, with the remaining low energy degrees of freedoms being the meson superfields. Therefore, the cutoff of the low energy meson theory is $v$, modulo order one corrections.

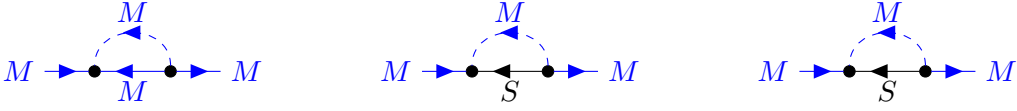
\begin{figure}[h]
  \centering
  \begin{subfigure}[t]{0.32\textwidth}
    \centering
    \feynmandiagram [layered layout, horizontal=a to d,scale = 0.8] {
  a[blue, particle = $M$] --[blue,fermion] b [dot] -- [blue,anti fermion,edge label' = $M$] c [dot] --[blue,fermion] d [blue, particle = $M$],
  c -- [blue, charged scalar, half right, looseness=1.5,edge label' = $M$] b,
 };
  \end{subfigure}
  \begin{subfigure}[t]{0.32\textwidth}
    \centering
    \feynmandiagram [layered layout, horizontal=a to d,scale = 0.8] {
  a[blue,particle = $M$] --[blue,fermion] b [dot] -- [anti fermion,edge label' = $S$] c [dot] --[blue,fermion] d [blue,particle = $M$],
  c -- [blue, charged scalar, half right, looseness=1.5,edge label' = $M$] b,
 }; 
\end{subfigure}
\begin{subfigure}[t]{0.32\textwidth}
    \centering
    \feynmandiagram [layered layout, horizontal=a to d,scale = 0.8] {
  a[blue,particle = $M$] --[blue,fermion] b [dot] -- [anti fermion,edge label' = $S$] c [dot] --[blue,fermion] d [blue,particle = $M$],
  c -- [blue, charged scalar, half right, looseness=1.5,edge label' = $M$] b,
 }; 
\end{subfigure}
\caption{Diagrams contributing to the mass correction of the adjoint.}
\label{fig:Adjoint_Mass_Correction_Diagrams}
\end{figure}

We calculate superfield renormalizations via fermion self-energy diagrams. The leading correction comes from the diagrams pictured below in Figure \ref{fig:Singlet_Mass_Correction_Diagrams} and \ref{fig:Adjoint_Mass_Correction_Diagrams}.

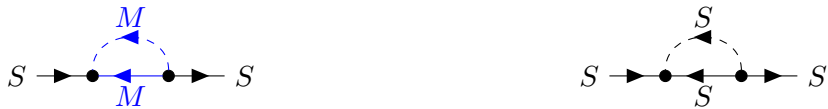
\begin{figure}[h]
  \centering
  \begin{subfigure}[t]{0.49\textwidth}
    \centering
    \feynmandiagram [layered layout, horizontal=a to d,scale = 0.8] {
  a[particle = $S$] --[fermion] b [dot] -- [blue,anti fermion,edge label' = $M$] c [dot] --[fermion] d [particle = $S$],
  c -- [blue, charged scalar, half right, looseness=1.5,edge label' = $M$] b,
 };
  \end{subfigure}
  \begin{subfigure}[t]{0.49\textwidth}
    \centering
    \feynmandiagram [layered layout, horizontal=a to d,scale = 0.8] {
  a[particle = $S$] --[fermion] b [dot] -- [anti fermion,edge label' = $S$] c [dot] --[fermion] d [particle = $S$],
  c -- [ charged scalar, half right, looseness=1.5,edge label' = $S$] b,
 };  
  \end{subfigure}
  \caption{Diagrams contributing to the mass correction of the singlet.}
  \label{fig:Singlet_Mass_Correction_Diagrams}
\end{figure}

The one-loop renormalization from any one particular diagram is of the form
\begin{align}
\delta Z_{1-loop} = -\frac{C \mu^2}{16\pi^2 v^2} \int_0^1 dz (1-z) \log \frac{-z(1-z) M_0^2 +(1-z) M_\text{bos}^2 + z M_\text{ferm}^2}{v^2}
\end{align}
Where $C$ is a Lagrangian and group theory dependent prefactor.

We must find the Yukawa couplings of the theory to find the values of the prefactors. We use Eq.~(\ref{kahler11}) to find the Christoffel symbols of the K\"{a}hler metric to be at leading order
\begin{align}
\Gamma^i_{jk} = -\frac{1}{2\sqrt{2}\,v} \Tr T_i \{ T_j,T_k \}
\end{align}
With this the Lagrangian's Yukawa term is
\begin{align}
\begin{split}
\mathcal{L}_\text{Yukawa} &= -\frac{1}{2} \left( \frac{\partial^2 W}{\partial \Mes_j \Mes_k} - \Gamma^i_{jk} \frac{\partial W}{\partial \Mes_i} \right) \psi_j \psi_k + h.c \\
&= \frac{1}{2} \frac{\sqrt{2}\, \mu}{v} \left( \frac{r(2r+1)}{\sqrt{N_f}} S \psi_S^2 + \frac{r+2}{\sqrt{N_f}} S \Tr \psi_M^2 + 2\frac{2r+1}{\sqrt{N_f}} \psi_S \Tr M\psi_M + 3\Tr M\psi_M^2 \right) + h.c.
\end{split}
\end{align}

The prefactors, dependent on the representation of the external, internal scalar, and internal fermion respectively, are found to be
\begin{gather}
C_{S,S,S} = \frac{2 r^2 (2r+1)^2}{N_f}\underset{r\rightarrow 2}{\longrightarrow} \frac{200}{N_f} \\
C_{S,M,M} = \frac{2 (N_f^2 - 1) (2r+1)^2}{N_f}\underset{r\rightarrow 2}{\longrightarrow} \frac{50(N_f^2-1)}{N_f}  \\
C_{M,M,M} = 9\frac{N_f^2 - 4}{N_f} \\
C_{M,S,M} = \frac{2 (r+2)^2}{N_f} \underset{r\rightarrow 2}{\longrightarrow} \frac{32}{N_f} \\
C_{M,M,S} = \frac{2 (2r+1)^2}{N_f}\underset{r\rightarrow 2}{\longrightarrow} \frac{50}{N_f} 
\end{gather}

Given that the K\"ahler potential and superpotential are expansions in powers of $v^{-1}$ while the cutoff is $v$, higher derivative operators can also contribute to the one-loop at $\mathcal{O}(\mu^2 / v^2)$. However, treating $m$ as a spurion of the explicitly broken chiral symmetry, one can show that these contributions come without the $\log(v/m)$ and are therefore subdominant.

Putting this all together gives the 1-loop correction to the correction to the mass ratio to be
\begin{align*}
     \delta_\text{1-loop} =&- \frac{(C_{SSS}+C_{SMM})-(C_{MMM}+C_{MSM}+C_{MMS})}{16\pi^2}\frac{\mu^2}{v^2}\log\left(\frac{v}{2\mu}\right) + \text{finite} \\
     =&-\frac{104+41 N_f^2 }{16\pi^2 N_f}\frac{\mu^2}{v^2}\log\left(\frac{v}{2\mu}\right) +\cdots
\end{align*}
Here we expanded in $\mu/v\ll1$, which is shown to be a consistent approximation in the region of parameter space fitting the self-interaction results of section \ref{section:Self Interactions}.

\section{Anomaly matching for neutral meson decay}\label{anom_match}

In this Appendix we discuss the conditions required to avoid neutral meson decay. We find the same condition as was found in a non-SUSY context by \cite{Berlin:2018tvf}. The decay of neutral mesons in the SRDM model parallels the decay of SM neutral pions. Even without the weak force, the decay $\pi^0 \rightarrow \gamma \gamma$ is not forbidden by $U(1)_{EM}$ charge conservation, and is in fact realized via anomalies. The process comes from the Wess-Zumino-Witten (WZW) term $~\pi^0 F \widetilde{F}$ in the pion effective Lagrangian. In the pion effective theory of the IR, a chiral rotation effects a shift $\pi^0 \rightarrow \pi^0 + \alpha$. Thus the WZW term ensures that the chiral anomaly produced by fermions (quarks) in the UV is matched in the IR where there are no fermions. The story in our model changes only slightly with the addition of SUSY where the IR mesons now have fermionic components.

We introduce a dark photon $U(1)_D$ that couples, in the UV theory, to the quark superfields with diagonal $N_f \times N_f$ charge matrix $Q$, which in the SM would be $\text{diag}(+\frac{2}{3},-\frac{1}{3})$. Maintaining a vector-like theory, the anti-quarks couple with matrix $-Q$. Let us perform an axial rotation using any diagonal generator $T$ of $SU(N_f)_A \times U(1)_R$ (note in the SM we have no $U(1)_R$ so that $T$ must be the traceless $\sigma^3$).\footnote{We exclude $U(1)_A$ because it carries the $U(1)_A SU(N_c)^2$ anomaly.} This acts on the quarks and anti-quarks of the UV theory. The $U(1)_T U(1)_D^2$ anomaly in the UV is $2 \Tr (T Q^2)$. The same quantity in the IR is
\begin{equation}
\begin{split}
\Tr_\text{M} T_\text{M} Q_\text{M}^2 & = \sum_{ij} (T_i + T_j)(Q_i - Q_j)^2 = 2 \sum_{ij} (T_i Q_i^2 -2 T_i Q_i Q_j + T_i Q_j^2) \\
 & = 2 (N_f \Tr (T Q^2) -2 \Tr (T Q) \Tr Q + \Tr T \Tr Q^2)
\end{split}
\end{equation}

The difference between the UV and IR anomalies must be made up by a WZW term that transforms at linear order as
\begin{equation}\label{wzw}
\sim 2 ((1-N_f) \Tr (T Q^2) +2 \Tr (T Q) \Tr Q - \Tr T \Tr Q^2) F_{\alpha \beta} \widetilde{F}^{\alpha \beta}
\end{equation}

The meson field without the VEV subtracted off, $\Mes'=v^2\mathbf{1}+\Mes$, transforms as $\Mes' \rightarrow e^{i T} \Mes' e^{i T} \approx \Mes' + 2i v^2 T$. Instead of calculating the full WZW term, we simply note that to reproduce the anomaly shift in Eq.~(\ref{wzw}), at linear order in $\Mes'$ we can have\footnote{Note that these coefficients do not necessarily match those of Eq.~(\ref{wzw}) because terms higher order in $\Mes'$ also transform at linear order in $T$. }
\begin{equation}\label{anom_cond}
\sim \left(c_1 \Tr (\Mes' Q^2) +c_2 \Tr (\Mes' Q) \Tr Q +c_3 \Tr \Mes' \Tr Q^2 \right) (W_\alpha W^\alpha |_{\theta^2})
\end{equation}
Generically these terms would give rise to decays of non-singlet neutral mesons, via off-shell dark photons, to SM particles. To avoid this, Eq.~(\ref{anom_cond}) should vanish. It is sufficient to require either that $Q \propto \mathbf{1}$ (mesons are uncharged under $U(1)_D$), or that $Q^2 \propto \mathbf{1}$ and $\Tr Q = 0$ as noted in \cite{Berlin:2018tvf}. We assume the latter condition holds throughout the paper.

Note that it may also be possible to come up with non-trivial charge configurations, depending on the coefficients $c_i$, that conspire to make the non-singlet terms of Eq.~(\ref{anom_cond}) disappear. However, for large enough $g_D$, the breaking of the quark degeneracy could be sizeable and could lead to the destruction of the resonance present in the meson spectrum.

\bibliographystyle{JHEP.bst}
\bibliography{bibliography.bib}

\end{document}